\documentclass[12pt]{article}

\usepackage{latexsym,amsmath,amssymb,theorem,epsfig}

\DeclareFontFamily{OT1}{rsfs10}{}
\DeclareFontShape{OT1}{rsfs10}{m}{n}{ <-> rsfs10 }{}
\DeclareMathAlphabet{\mathscript}{OT1}{rsfs10}{m}{n}

\numberwithin{equation}{section}

\newcommand{\ns}{\normalsize}

\newcommand{\tr}{\text{tr}}

\def\b{\beta}

\def\e{\epsilon}

\theoremstyle{plain}

{\theorembodyfont{\rmfamily} }

\begin{document}


\begin{titlepage}

\vspace{-5cm}

\title{
  \hfill{\ns }  \\[1em]
   {\LARGE Raising Anti de Sitter Vacua to de Sitter Vacua in Heterotic M-Theory}
\\[1em] }
\author{
   Evgeny I. Buchbinder
     \\[0.5em]
   {\ns School of Natural Sciences, Institute for Advanced Study} \\[-0.4cm]
{\ns Einstein Drive, Princeton, NJ 08540}\\[0.3cm]}

\date{}

\maketitle

\begin{abstract}

We explore the possibility of obtaining de Sitter vacua in
strongly coupled heterotic models by adding various corrections
to the supergravity potential energy. We show that, in a generic
compactification scenario, Fayet-Iliopoulos terms can generate a
de Sitter vacuum. The cosmological constant in this vacuum can be
fine tuned to be consistent with observation. We also study
moduli potentials in non-supersymmetric compactifications of $E_8
\times E_8$ theory with anti five-branes and $E_8 \times \bar
E_8$ theory. We argue that they can be used to create a de Sitter
vacuum only if some of the Kahler structure moduli are stabilized
at values much less than the Calabi-Yau scale.

\end{abstract}

\thispagestyle{empty}

\end{titlepage}


\section{Introduction}


The moduli stabilization problem is one of the central in search
for realistic string theory vacua in four dimensions. On one
hand, the existence of massless scalar fields is in conflict with
experiment. On the other, the four-dimensional gravitational and
gauge coupling constants depend on the values of the moduli.
Therefore, in any realistic string compactification, all moduli
have to be stabilized in a phenomenologically acceptable range.
Another four-dimensional quantity, apparently intrinsically
related to the moduli stabilization problem, is the cosmological
constant. Recently, substantial progress in this direction was
achieved in the work of Kachru, Kallosh, Linde and Trivedy
\cite{KKLT}. In the context of Type IIB flux compactifications
\cite{DRS, GKP, Schulz1, KSTT, FP, GVW, Taylor, Curio,
Gir}\footnote{Flux compactifications were also found promising
for moduli stabilization in M-theory on singular $G_2$ manifolds
in~\cite{Bobby}}, in \cite{KKLT}, it was shown that it is
possible to stabilize all moduli in a metastable de Sitter (dS)
vacuum. The stabilization procedure in \cite{KKLT} was performed
in two steps. At step one, all moduli were stabilized in an anti
de Sitter (AdS) minimum. This was done by balancing fluxes
against non-perturbative effects \cite{DSWW1, DSWW2, BBS,
Witten95}. At step two, it was demonstrated that the minimum can
be lifted to a metastable dS vacuum, by adding anti D3-branes to
the system. A crucial ingredient at this step was the result of
\cite{KPV} that a flux-anti D3-brane system can form a metastable
bound state with positive energy. The effect of combining this
positive contribution to the potential energy with the negative
supergravity contribution can produce a dS vacuum. Furthermore,
in \cite{KKLT}, it was shown that it is possible to fine tune the
cosmological constant and make it consistent with observations.
Later, moduli stabilization in this Type IIB scenario was
explored in more detail in \cite{Silver, Sethi, Douglas}.

It is a natural question whether a similar moduli stabilization procedure can be
fulfilled in a more realistic framework of strongly coupled heterotic string theory, or,
heterotic M-theory \cite{HorWit1, HorWit2, Witten96}.
Such compactifications have a lot of phenomenologically attractive features
(see \cite{Faraggi} for a recent review on phenomenological aspects of M-theory).
Various GUT- and Standard Model-like theories were obtained from
heterotic compactifications on a Calabi-Yau manifold \cite{DLOW, DOPW, DOPR}.
The actual particle spectrum in such theories was recently studied in \cite{Yang1, Yang2}.
Progress towards construction of dS vacua from heterotic M-theory was recently
reported in \cite{BCK}\footnote{Questions concerning under what circumstances a dS vacuum can arise in
heterotic string theory were also studied in~\cite{DeAl}},
based on the earlier work \cite{CurK}. A dS vacuum was obtained
by balancing two non-perturbative effects, gaugino condensate \cite{DRSW, Horava, LOW}
and open membrane instantons \cite{Lima1, Lima2, Saulina} as well as
by using perturbative potentials for charged matter fields. Despite obvious progress,
this method has certain shortcomings. First, not all moduli have been stabilized.
In particular, complex structure moduli remained unfixed. In order to stabilize them,
apparently, it is necessary to introduce flux-induced superpotentials \cite{GVW, Berndt, Constantin}.
However, results from \cite{KKLT, BO, GKLM} suggest that that such superpotentials
tend to stabilize moduli in AdS minima. Second, authors in \cite{BCK} gave
vacuum expectation values (vevs) to the charged matter fields. This
breaks the low-energy gauge group down to nothing unless the vevs are under control.
One more problem is that it seems hard to stabilize all run away moduli this way.
This, though, might be avoided by using non-Kahler background
\cite{Becker1, Becker2, Becker3, Gott1, Gott2, Cardoso}. It would be great to overcome
these problems because in \cite{BCK} no fine tuning was required
to produce a phenomenologically attractive dS vacuum.

In any case, it seems important to create a heterotic analogue of
\cite{KKLT}. The first step associated with stabilizing moduli in
an AdS minimum was proposed in \cite{BO} in strongly coupled
heterotic string theory and in \cite{GKLM} in the weakly coupled
case. In \cite{BO}, it was shown in a very general set-up that all
heterotic moduli, including complex and Kahler structure moduli,
the volume modulus, vector bundle moduli and five-brane moduli,
could be stabilized in an AdS vacuum in a phenomenologically
acceptable range. In this paper, we would like to explore the
possibility of performing the second step of \cite{KKLT} and
lifting this AdS minimum to a dS minimum. To be more precise, we
consider two different ways how such a lift can be achieved.
First, we study in detail moduli potentials induced by
Fayet-Iliopoulos terms \cite{DSW}. Such terms arise when the
low-energy field theory contains an anomalous $U(1)$ factor. The
anomaly cancels by the Green-Schwarz mechanism. The idea that
such potentials can produce a dS minimum was suggested by
Burgess, Kallosh and Quevedo in \cite{Burgess}. We apply this
idea for the case of strongly coupled heterotic string theory and
argue that, generically, the order of magnitude of such
potentials is small enough, so that it is possible to balance
them against fluxes and non-perturbative effects. Then we show
that it is indeed possible to obtain a metastable dS vacuum along
these lines. At this point, we should note that Fayet-Iliopoulos
terms modify the potential for charged matter fields as well and
might be important for understanding issues related to the
supersymmetry breaking scale. In this paper, we will not address
these questions. Matter potentials represent an independent
difficult problem which requires a serious study. The goal of
this paper is to show that it is possible to stabilize all
heterotic string moduli in a dS vacuum.

The second method to raise AdS minima that we consider is more universal as
it does not put any restrictions on the structure of the low-energy physics.
This method is based on adding anti five-branes in the bulk. A
more general set-up would be to consider the $E_8 \times \bar
E_8$ theory with (anti) five-branes present in the bulk. The $E_8
\times \bar E_8$ theory was introduced by Fabinger and Horava
in~\cite{Fabinger}. It is obtained by a chirality flip at one of
the orbifold planes. This is equivalent to having supergravity on
$S^1/Z_2$ with the gravitino antiperiodic along the circle. This
imples that in the effective field theory, the gravitino will
have a mass, whereas in the matter sector, supersymmetry will be
preserved, at least at the tree level. In \cite{Fabinger}, it was
argued that the orbifold fixed planes experince the attractive
Casimir-type force. Upon compactifying this theory to
four-dimensions on a Calabi-Yau manifold, this effect becomes
subleading. The leading potential is induced by non-trivial
charges, depending on the second Chern classes of the gauge and
the tangent bundles, on these planes. Similar potentials arise if
one adds (anti) five-branes in either $E_8 \times E_8$ or $E_8
\times \bar E_8$ theory. However, we obtain that, for a generic
compactification, the order of magnitude of this new potential is
too big comparing to the order of magnitude of fluxes and
non-perturbative effects. Therefore, it destabilizes the vacuum
rather than just modifying it. A way to resolve this problem can
be to take a Calabi-Yau which has cycles of various sizes. Then
it is possible to decrease this new correction to the potential
energy. To do this, it is necessary to stabilize $h^{1,1}$ moduli
of the Calabi-Yau manifold in such a way that various cycles are
fixed at different scales. This problem seems to be related to
understanding non-exponential factors in non-perturbative
superpoptentials \cite{BDO1, BDO2} and will not be discussed in
this paper.

This paper is organized as follows. In Section 2, we consider a
system having an AdS minimum. This system involves the volume
modulus, the interval modulus, one five-brane modulus and the
complex structure moduli. This is a simplified version of the one
studied in \cite{BO}. In \cite{BO}, it was shown that the
remaining moduli of the vector bundle can be stabilized as well.
It was also argued in \cite{BO} that any number of $h^{1,1}$
moduli can be stabilized by similar mechanism. The necessity of a
five-brane is dictated by the fact that the non-perturbative
superpotential for the interval modulus dies off too fast.
Therefore, it is problematic to stabilize it. However, if the
five-brane is located close enough to one of the orbifold fixed
plane one can stabilize the interval modulus. The complex
structure moduli are stabilized by the flux-induced
superpotential, two out of three remaining moduli are stabilized
by balancing run away moduli against the fluxes. For the last
modulus, we obtain a purely algebraic equation. A numeric
analysis shows that it is possible to find a solution satisfying
all requirements and assumptions. In Section 3, we discuss the
contribution to the potential energy induced by Fayet-Iliopoulos
terms \cite{DSW, Burgess}. The form of the potential is slightly
different depending on in what sector there is an anomalous
$U(1)$ gauge group. This contribution depends on the volume as
well as on the interval and the five-brane moduli. We estimate
its order of magnitude and find that, for a generic
compactification, it is comparable with that of fluxes and
non-perturbative effects. In Section 4, we consider the system
from Section 2 and modify the potential energy by a
Fayet-Iliopoulos term. By explicit calculation, we show that it is
indeed possible to produce a dS minimum. We also show that it is
also conceivable to obtain a small cosmological constant by fine
tuning. The reason for this is that the supergravity contribution
to the cosmological constant can still be kept negative. However,
we note that if the number of moduli is large enough it is very
hard to keep the supergravity potential energy negative in the
vacuum. As a consequence, in this case, it is still possible to
find a dS minimum but the cosmological constant will always be
very large. Even though we do not include vector bundle moduli
into our analysis, we comment that it is straightforward to add
them without facing conceptual difficulties. In Section 5, we
move on to the $E_8 \times \bar E_8$ theory. By simple anomaly
arguments along the lines of \cite{HorWit2}, we find a necessary
condition for anomaly free compactifications. Then, we derive the
moduli potential in this theory. We show that a potential with
the identical functional structure arises from adding (anti)
five-branes. This is, of course, not surprising. In a
supersymmetric $E_8 \times  E_8$ compactification without anti
five-branes, the potential is identically zero as the consequence
of the net tension cancellation. It is interesting to note that
this potential can be both positive and negative. It depends on
the Calabi-Yau volume modulus and on the interval modulus. Even
though we do not do explicit calculations with this potential, it
is natural to expect that it works as good as Fayet-Iliopoulos
terms. We estimate its order of magnitude and find that, for a
generic compactification, it is not comparable with that of
fluxes and non-perturbative effects. We give a brief discussion
on what it takes to decrease the order of magnitude of this
potential. The key issue seems to be to learn how to stabilize
some of the Kahler structure moduli of a Calabi-Yau manifold at
scales sufficiently smaller than the Calabi-Yau scale. This
method to introduce a correction to the supergravity Lagrangian
is, in a certain sense, more attractive than to use
Fayet-Iliopoulos terms. First, it is more universal. It does not
impose any constraints on the structure of the low-energy gauge
group. Second, this new contribution to the potential energy can
be both positive and negative. This might be important for the
purposes of fine tuning the cosmological constant.

In this paper, we work in the framework of Calabi-Yau compactifications.
It is natural to
expect that similar results should hold in the context
of non-Kahler compactifications
\cite{Becker1, Becker2, Becker3, Gott1, Gott2, Cardoso}.
In such compactifications, the volume of the manifold is stabilized perturbatively.
Therefore, the analysis can be very similar from the conceptual viewpoint but simpler technically
since a fewer number of moduli is involved.
However,
it is hard to say exactly what the structure of Kahler potentials and superpotentials is,
because the moduli of non-Kahler compactifications are not known.
In particular, since there are no $h^{1,1}$ moduli, the structure of non-perturbative superpotentials
is unclear. These subtleties require detailed investigations.


\section{AdS Vacua}


As in \cite{KKLT}, we begin with the construction of the AdS minimum. This was done in a
very general setting in \cite{BO}. Here we consider a simplified system where
we ignore vector bundle moduli.
The details of their stabilization can be found in \cite{BO}.
We work in the context of the strongly coupled heterotic string theory~\cite{HorWit1, HorWit2}.
To one of the orbifold fixed planes we will refer as to the visible brane (or the visible sector),
to the other one we will refer as to the hidden brane (or the hidden sector).
The system under study involves the following complex moduli
\begin{equation}
S, T, {\bf Y}, Z_{\alpha} .
\label{1.1}
\end{equation}
The modulus $S$ is related to the volume of the Calabi-Yau manifold
\begin{equation}
S=V+i \sigma ,
\label{1.2}
\end{equation}
where $\sigma$ is the axion. The real part of the $T$-modulus is the size of the eleventh dimension
\begin{equation}
T=R+i p,
\label{1.3}
\end{equation}
where $p$ comes from the components of the M-theory three-form $C$ along the interval and the
Calabi-Yau manifold. Here we have assumed that $h^{1,1}=1$. In \cite{BO}, it was argued that
one should be able to stabilize any number of $h^{1,1}$ moduli by solving similar but technically
more complicated equations. As in \cite{BO}, we will do all calculations assuming that there is
only one $h^{1,1}$ modulus which we denote by $T$.
{\bf Y} is the modulus of the five-brane. Throughout the paper, we assume
that there is only one five-brane in the bulk wrapping an isolated genus zero curve. In this case,
there is only one five-brane modulus \cite{Five}, whose real part is the position of the five-brane
in the bulk
\begin{equation}
{\bf Y}=y+i(a+\frac{p}{R}) ,
\label{1.4}
\end{equation}
where $a$ is the axion arising from dualizing the three-form field strength propagating on the
five-brane world-volume. At last, by $Z_{\alpha}$ we denote the complex structure moduli. The actual
number of them is not relevant for us. The moduli $V, R$ and $y$ are assumed to be dimensionless
normalized with respect to the following reference scales
\begin{equation}
v_{CY}^{-1/6} \approx 10^{16} GeV, \quad (\pi \rho)^{-1} \approx 10^{14}-10^{15} GeV.
\label{1.4.1}
\end{equation}
In order to obtain the four-dimensional coupling constants in the correct phenomenological range
\cite{Witten96, Banks},
the corresponding moduli should be stabilized at (or be slowly rolling near) the values
\begin{equation}
V \sim 1 \quad R \sim 1.
\label{1.4.2}
\end{equation}

The Kahler potential for this system is as follows \cite{Candelas, LOW4, DerS}
\begin{equation}
\frac{K}{M^{2}_{Pl}} = K_Z + K_{S,T,{\bf Y}},
\label{1.5}
\end{equation}
where
\begin{equation}
K_Z= -\ln(-i \int \Omega \wedge \bar \Omega),
\label{1.5.1}
\end{equation}
and
\begin{equation}
K_{S,T,{\bf Y}} =-\ln(S+\bar S) -3 \ln (T+\bar T) +2 \tau_5 \frac{({\bf Y}+\bar{\bf Y})^2}{(S+\bar S)(T+\bar T)}.
\label{1.5.2}
\end{equation}
Here $M_{Pl}$ is the four-dimensional Planck scale and $\tau_5$ is given by
\begin{equation}
\tau_5 =\frac{T_5 v_5 (\pi \rho)^2}{M^{2}_{Pl}},
\label{1.5.3}
\end{equation}
where $v_5$ is the area of the cycle on which the five-brane is wrapped and $T_5$ is
\begin{equation}
T_5 =(2 \pi)^{1/3} (\frac{1}{2 \kappa_{11}^2})^{2/3},
\label{1.5.34}
\end{equation}
with $\kappa_{11}$ being the eleven-dimensional gravitational coupling constant. It is
related to the four-dimensional Planck
mass as
\begin{equation}
\kappa_{11}^2=\frac{\pi \rho v_{CY}}{M^2_{Pl}}.
\label{1.5.4}
\end{equation}
Evaluating $\tau_5$ by using \eqref{1.5.4} and \eqref{1.4.1} gives
\begin{equation}
\tau_5 \approx \frac{v_5}{v_{CY}^{1/3}}.
\label{1.5.5}
\end{equation}
Generically this coefficient is of order one.

The superpotential for this system consists of three different contributions
\begin{equation}
W=W_{f}-W_{g} -W_{np}.
\label{1.5.6}
\end{equation}
$W_f$ is the flux-induced superpotential \cite{GVW, Berndt, Constantin}
\begin{equation}
W_f =\frac{M^2_{Pl}}{v_{CY}} \int_{CY} H \wedge \Omega,
\label{1.6}
\end{equation}
where $H$ is the Neveu-Schwarz three form. In M-theory notation it can be written as
\begin{equation}
W_f =\frac{M^2_{Pl}}{v_{CY}\pi \rho} \int dx^{11} \int_{CY} G \wedge \Omega,
\label{1.6.1}
\end{equation}
where $G$ is the M-theory four-form flux. The order of magnitude
of $W_f$ was estimated in \cite{BO} and was found to be,
generically, of order $10^{-8}M_{Pl}^3$. In fact, this is
flexible. The superpotential $W_f$ may receive certain higher
order corrections from Chern-Simons invariants. In \cite{GKLM} it
was argued that this Chern-Simons invariants can reduce the order
of magnitude of $W_f$. We will assume in this paper that the
order of magnitude of $W_f$ is approximately $10^{-10}-10^{-9}$
in Planck units. It is well known that perturbatively a $(3,0)$
Neveu-Schwarz flux (or, equivalently, a $(3,0,1)$ $G$-flux in
M-theory) breaks supersymmetry. Therefore, the idea is to balance
the superpotential \eqref{1.6} against the non-perturbative
contributions $W_g$ and $W_{np}$. At this point, it is appropriate
to mention that we are not allowing $(2,1,1)$ components of the
M-theory $G$-flux which leads to non-Kahler compactifications
\cite{Becker1, Becker2, Becker3, Gott1, Gott2, Cardoso}. The flux
introduced above provides only a perturbative deformation of the
Calabi-Yau metric.

By $W_g$ we denote the superpotential induced by a gaugino
condensate in the hidden sector \cite{DRSW, Horava, LOW,
Nonstandard}. A non-vanishing gaugino condensate has important
phenomenological consequences. Among other things, it is
responsible for supersymmetry breaking in the hidden sector. When
that symmetry breaking is transported to the observable brane, it
leads to soft supersymmetry breaking terms for the gravitino,
gaugino and matter fields \cite{KL, BIM, NOY, LT}. See~\cite{Nil}
for a good review on gaugino condensation in string theory. This
superpotential has the following structure
\begin{equation}
W_{g} = h M^{3}_{Pl} exp(-\e S +\e \alpha^{(2)} T - \e \b \frac{{\bf Y}^2}{T}).
\label{1.7}
\end{equation}
The order of magnitude of $h$ is approximately $10^{-6}$ \cite{LOW}. The coefficient $\e$ is related to the
coefficient $b_0$ of the one-loop beta-function and is given by
\begin{equation}
\e = \frac{6 \pi}{b_0 \alpha_{GUT}}.
\label{1.7.1}
\end{equation}
For example, for the $E_8$ gauge group $\e \approx 5$.
The coefficient $\alpha^{(2)}$ represents the
tension (up to the minus sign) of the hidden brane
\begin{equation}
\alpha^{(2)} \sim
\frac{\pi \rho}{16 \pi v_{CY}}
(\frac{\kappa_{11}}{4 \pi})^{2/3}
 \int_{CY} \omega \wedge (tr F^{(2)} \wedge F^{(2)} - \frac{1}{2} tr {\cal R}
\wedge {\cal R}),
\label{1.7.2}
\end{equation}
where $\omega$ is the Kahler form and $F^{(2)}$ is the curvature of the gauge bundle on the hidden brane.
Similarly, the coefficient $\beta$ is the tension of the five-brane. It is given by \cite{Visible}
\begin{equation}
\beta=\frac{2 \pi^2 \rho}{v_{CY}^{2/3}} (\frac{\kappa_{11}}{4 \pi})^{2/3}
\int_{CY} \omega_{I} \wedge {\cal W},
\label{1.7.3}
\end{equation}
where ${\cal W}$ is the four-form Poincare dual to the holomorphic curve on which
the five-brane is wrapped.
Generically both $\alpha^{(2)}$ and $\beta$ are of order one.
In fact, from eqs.~\eqref{1.5.3}, \eqref{1.5.34} and and \eqref{1.7.3} it follows that
\begin{equation}
\b \approx \tau_5.
\label{1.7.4}
\end{equation}
The quantity
\begin{equation}
Re(S - \alpha^{(2)} T + \b \frac{{\bf Y}^2}{T})
\label{1.8}
\end{equation}
represents the inverse square of the gauge coupling constant in
the hidden sector, $\frac{1}{g^{2}_{hidden}}$. Therefore, it
cannot become negative. This, in particular, says that the
superpotential~\eqref{1.7} cannot be trusted for large values of
the interval length $R$. We believe that higher order corrections
to the combination \eqref{1.8}  will make the gauge coupling
constant well defined for large values of $R$. Partial support
for this comes from the work of Curio and Krause \cite{Curio1,
Curio2} who showed that the next order order $T$-correction to
$\frac{1}{g^{2}_{hidden}}$ is indeed positive. The difficulty
with understanding $W_g$ for sufficiently large values of $R$
leads to necessity of introducing five-branes and non-perturbative
superpotentials. If we ignore the ${\bf Y}$-modulus and restrict
ourselves to the $S, T, Z_{\alpha}$-system with the superpotential
\begin{equation}
W_{f}-W_{g},
\label{1.8.1}
\end{equation}
it is straightforward to show that the potential energy is strictly positive definite unless
\begin{equation}
\frac{1}{g^2_{hidden}} < 0.
\label{1.8.2}
\end{equation}
This, apparently, implies that we cannot trust $W_{g}$ in the form \eqref{1.7} and have to include
higher order corrections. Instead of doing that, we will add the five-brane and show that it is
possible to find an interesting solution without running into troubles
with the imaginary gauge coupling constant.
One more problem with the superpotential \eqref{1.8.1} is that it can stabilize
only one linear combination of the imaginary parts of $S$ and $T$ moduli leaving the remaining one flat.
In order to be able to balance $W_f$ and $W_g$ with each other,
their orders of magnitude should approximately be the same.
This is clearly possible, especially with a help of Chern-Simons invariants,
if $V$ and $R$ and of order one and $V-\alpha^{(2)}R$ is positive.

The last contribution to the superpotential that we have to discuss is the non-perturbative superpotential
$W_{np}$
\cite{DSWW1, DSWW2, Witten95, BBS, Witten00, Lima1, Lima2, BDO1, BDO2}.
In principle, it has three parts
\begin{equation}
W_{np}=W_{vh}+W_{v5} +W_{5h}.
\label{1.9}
\end{equation}
$W_{vh}$ is induced by a membrane stretched between the visible and the hidden branes. It behaves as
\begin{equation}
W_{vh} \sim e^{-\tau T}
\label{1.10}
\end{equation}
$W_{v5}$ is induced by a membrane stretched between the visible brane and the five-brane. It behaves as
\begin{equation}
W_{v5} \sim e^{-\tau {\bf Y}}.
\label{1.11}
\end{equation}
At last, $W_{5h}$ is induced by a membrane stretched between the five-brane and the hidden brane. It behaves as
\begin{equation}
W_{5h} \sim e^{-\tau (T-{\bf Y})}.
\label{1.12}
\end{equation}
The coefficient $\tau$ is given by~\cite{Lima1, Lima2}
\begin{equation}
\tau =\frac{1}{2} (\pi \rho) v_z (\frac{\pi}{2 \kappa_{11}})^{1/3},
\label{1.13}
\end{equation}
where $v_z$ is the area of the holomorphic curve. Taking $v_z \approx v_{CY}^{1/3}$ and using~\eqref{1.4.1}
and \eqref{1.5.4}, we obtain
\begin{equation}
\tau \approx 250.
\label{1.14}
\end{equation}
Since we are interested in the regime $Re(T) \sim 1$ it is very difficult to make $W_{np}$ of order $W_f$.
The only situation when these two contributions to the superpotential can compete is when the five-brane
is close to one of the orbifold fixed planes. If the five-brane is close to the visible brane then
both $W_{vh}$ and $W_{5h}$ will die off very fast and the non-perturbative superpotential will not
depend on $T$. As a consequence, the $T$-modulus cannot be stabilized. Therefore, the only way to proceed
is to assume that the five-brane is closed to the hidden brane. In this case
\begin{equation}
W_{np} = W_{5h} = M^{3}_{Pl} a e^{-\tau (T-{\bf Y})}.
\label{1.15}
\end{equation}
For concreteness we assume that the coefficient $a \sim 1$. We will demand that
\begin{equation}
\tau e^{-\tau (T-{\bf Y})} \sim W_f \sim 10^{-10}.
\label{1.16}
\end{equation}
This says that
\begin{equation}
Re(T - {\bf Y}) \approx 0.1.
\label{1.17}
\end{equation}
Thus, we will assume that the five-brane is close to the hidden brane and take $W_{np}$ to be given by \eqref{1.15}.
Of course, we have to show that it is possible to stabilize the five-brane at such a distance.

Now we show following~\cite{BO} that the system under consideration indeed has an AdS minimum satisfying
\begin{equation}
D_{all {\ } fields} W=0,
\label{1.17.1}
\end{equation}
where $D$ is the Kahler covariant derivative,
and all
the assumptions stated above. For simplicity, we will look only at the real parts of the moduli. All imaginary parts
can be stabilized as well.
See \cite{BO} for details. Furthermore, we will not distinguish between the superpotentials and their absolute values.
First, we look at the equation
\begin{equation}
D_{Z_{\alpha}}W=0.
\label{1.18}
\end{equation}
Assuming that
\begin{equation}
W_{f} >> W_{g}, W_{np}
\label{1.19}
\end{equation}
in the interesting regime, eq.~\eqref{1.18} can be written as
\begin{equation}
\partial_{Z_{\alpha}}W_{f} + \frac{\partial K_{Z_{\alpha}}}{\partial Z_{\alpha}} W_{f} =0.
\label{1.20}
\end{equation}
In \cite{BO} it was shown that eq.~\eqref{1.19} is indeed satisfied. In eq.~\eqref{1.20},
all quantities depend on the complex structure moduli only. We will assume that this equation
fixes all the complex structure moduli. Partial evidence that equations of the type \eqref{1.20}
fix all the complex structure moduli comes, for example, from~\cite{Schulz1}.
The next equation to consider is
\begin{equation}
D_{S}W=0.
\label{1.20.1}
\end{equation}
By using eqs.~\eqref{1.5.2}, \eqref{1.7} and \eqref{1.19} we can rewrite this as
\begin{equation}
2\e V e^{-\e V + \dots} = (1+ \frac{2 \tau_5 y^2}{VR}) W_f.
\label{1.21}
\end{equation}
Eq.~\eqref{1.21} provides stabilization of the volume $V$. It is conceivable to find a
solution
\begin{equation}
V \sim 1.
\label{1.22}
\end{equation}
By using eqs.~\eqref{1.5.2}, \eqref{1.7}, \eqref{1.15}, \eqref{1.9} and \eqref{1.21},
eq.
\begin{equation}
D_T W=0
\label{1.22.1}
\end{equation}
can be rewritten as
\begin{equation}
\tau W_{np} = (\frac{3}{2R} +\tau_5 \frac{y^2}{VR^2}+ \frac{1}{2V}(\alpha^{(2)}+
\frac{\b y^2}{R^2})(1+\frac{2 \tau_5 y^2}{VR}))W_f.
\label{1.23}
\end{equation}
The left hand side of this equation is
\begin{equation}
\tau e^{-\tau (R-y)}
\label{1.23.1}
\end{equation}
The right hand side is of order $W_f$. As discussed before, this implies that
\begin{equation}
R-y=0.1.
\label{1.24}
\end{equation}
If it is possible to make the flux-induced superpotential $W_f$ smaller, then the difference $R-y$ can be increased.
Eq.~\eqref{1.24} stabilizes $R$ provided we can stabilize the five-brane. By using eqs.~\eqref{1.5.2}, \eqref{1.7},
\eqref{1.15}, \eqref{1.21} and \eqref{1.23}, eq.
\begin{equation}
D_{{\bf Y}}W=0
\label{1.25}
\end{equation}
can be reduced the following purely algebraic equation
\begin{equation}
(1+\frac{2\tau_5 y^2}{VR})(\frac{\b y}{R}-\frac{\alpha^{(2)}}{2}-\frac{\b y^2}{2 R^2})-\frac{3}{2} -\frac{\tau_5 y^2}{R^2}
+ \frac{2 \tau_5 y}{R} =0.
\label{1.26}
\end{equation}
A numeric analysis shows that it possible to find a solution for $y$ satisfying \eqref{1.24}
when $V$ and $R$ are both or order one.
For example, if we take
\begin{equation}
\alpha^{(2)} =1, \quad \b =1.5, \quad \tau_5 = 1.5, \quad V=1.2, \quad R=0.8,
\label{1.27}
\end{equation}
then eq.~\eqref{1.26} has a unique positive solution for $y$ given by
\begin{equation}
y \approx 0.7.
\label{1.28}
\end{equation}
Thus, we have shown that the system \eqref{1.17.1} can have a
solution satisfying all our assumptions and requirements. It is
easy to see that the combination~\eqref{1.8} is positive and of
order one.

The value of the potential energy in the vacuum is
\begin{equation}
U_{min} \sim - W_f^2.
\label{1.29}
\end{equation}
Our goal will be to raise this vacuum to a metastable dS vacuum.
First, we consider to the correction to the potential energy due
to Fayet-Iliopoulos terms. Then we will take a look at the $E_8
\times \bar E_8$ theory.

\begin{figure}
\epsfxsize=3.7in \epsffile{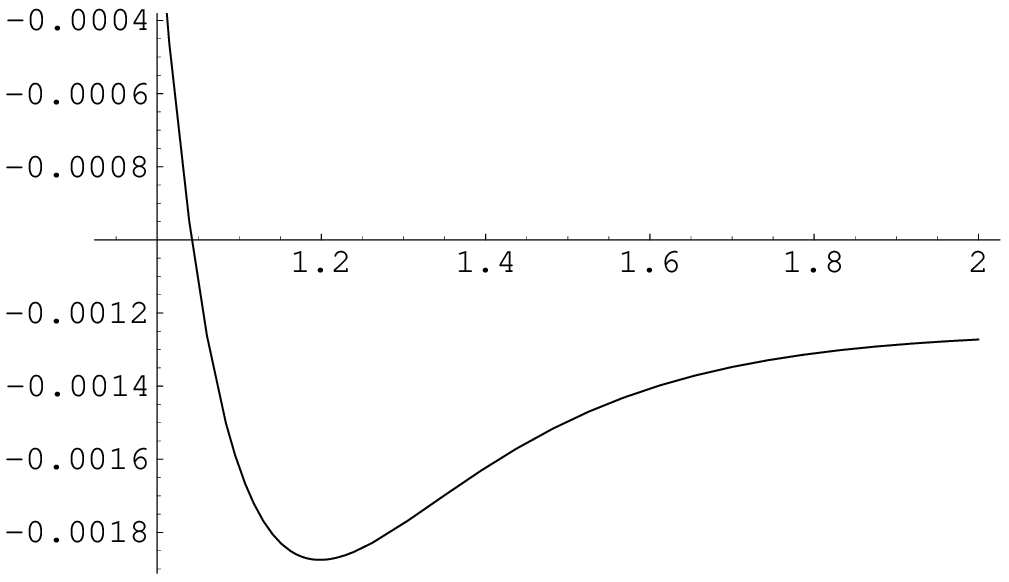}
\begin{picture}(30,30)
\put(-220,150){U}
\put(10, 85){V}
\end{picture}
\caption{A schematic slice of the potential near the AdS minimum
(multiplied by $10^{12}$) in the $V$ direction. \label{v}}
\vspace{0.5cm}
\epsfxsize=3.7in \epsffile{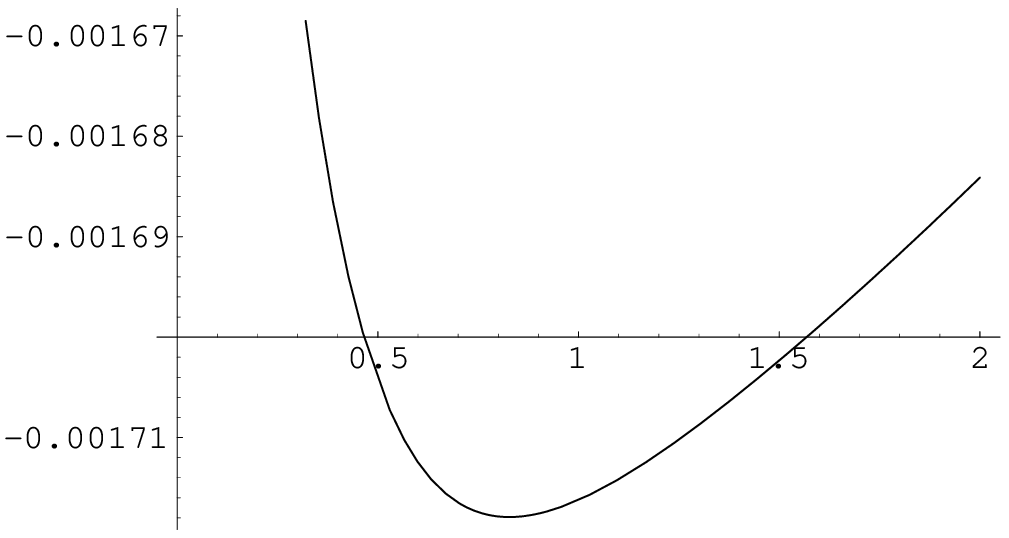}
\begin{picture}(30,30)
\put(-220,135){U}
\put(10, 55){R}
\end{picture}
\caption{A schematic slice of the potential near the AdS minimum
(multiplied by $10^{12}$) in the $R$ direction. \label{r}}
\vspace{0.5cm}
\epsfxsize=3.7in \epsffile{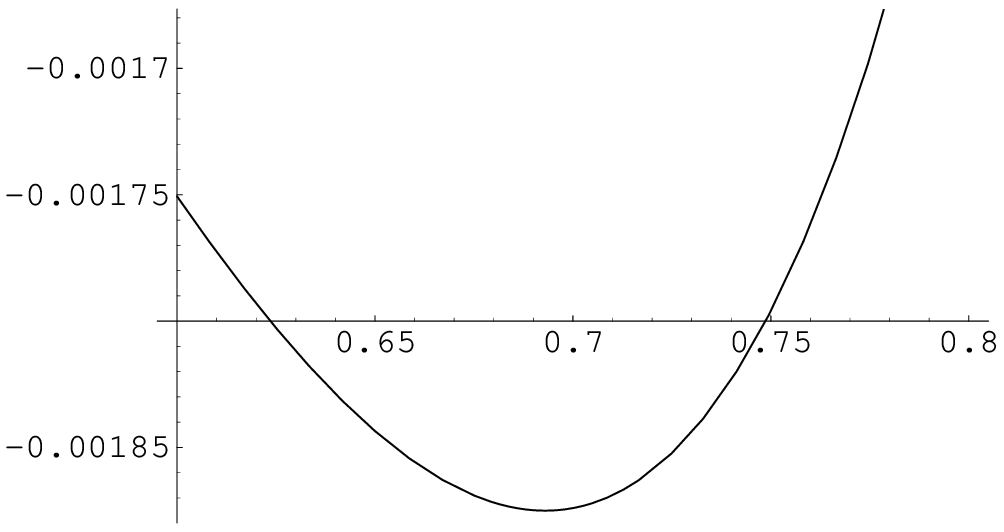}
\begin{picture}(30,30)
\put(-220,135){U}
\put(10, 55){y}
\end{picture}
\caption{A schematic slice of the potential near the AdS minimum
(multiplied by $10^{12}$) in the $y$ direction. \label{y}}
\end{figure}

Before we conclude this section, let us make some comments on the
imaginary parts and of the $S, T$ and ${\bf Y}$ moduli. Details
can be found in \cite{BO}. Imaginary parts are stabilized in such
a way that the superpotentials $W_g$ and $W_{np}$ are out of phase
with respect to $W_f$. We have already indicated this fact in
eq.~\eqref{1.5.6} by choosing the appropriate minus signs. The
imaginary part of the $T$-modulus behaves as
\begin{equation}
Im(T) \sim \frac{1}{\tau} \approx 0
\label{1.30}
\end{equation}
since $\tau$ is very large. One can also show that
\begin{equation}
Im ({\bf Y}) \approx 0.
\label{1.31}
\end{equation}
Various slices of the potential energy are schematically shown on Figures~\ref{v}-\ref{y}.
%


\section{Moduli Potentials From Fayet-Iliopoulos Terms}


In both weakly and strongly coupled heterotic string models there can be anomalous $U(1)$ gauge groups.
They can arise in both the visible and the hidden sectors. The anomaly is cancelled by the
four-dimensional version of the Green-Schwarz mechanism. To cancel the anomaly the axion $\sigma$
must undergo the gauge transformation of the form
\begin{equation}
\sigma \rightarrow \sigma + c \lambda,
\label{2.1}
\end{equation}
where $\lambda$ is a parameter and $c$ is a constant whose order
of magnitude will be estimated later. The transformation
law~\eqref{2.1} implies \cite{DSW} that the Kahler potential for
the $S$-modulus has to be modified as follows
\begin{equation}
K_{S} =-M^{2}_{Pl} \ln(S +\bar S + c {\cal V}),
\label{2.2}
\end{equation}
where ${\cal V}$ is the anomalous $U(1)$ vector superfield. From here we find that the kinetic term of the action
\begin{equation}
\int d^4x d^4 \theta K_S
\label{2.3}
\end{equation}
contains among others the Fayet-Iliopoulos term
\begin{equation}
\int d^4x M^{2}_{Pl} \frac{c}{S+\bar S} D,
\label{2.5}
\end{equation}
where $D$ is the auxiliary field of the vector multiplet. This gives rise to the moduli potential
energy of the form
\begin{equation}
U_{D} \sim M^{4}_{Pl} g^2 \frac{c^2}{V^2},
\label{2.6}
\end{equation}
where $g$ is the gauge coupling which is itself moduli dependent.
Note that $U_{D}$ is strictly positive.
Since the gauge coupling constants
are different in the visible and in the hidden sectors, the precise form of the
potential energy $U_D$ depends on in which sector there appeared an anomalous $U(1)$ gauge group.
The gauge coupling constants in the visible and the hidden sectors are given by \cite{Nonstandard}
\begin{equation}
g_{visible}^2 =\frac{g^2_0}{Re(S+\alpha^{(1)}T +\b (T-\frac{{\bf Y}^2}{T}))}
\label{2.6.1}
\end{equation}
and
\begin{equation}
g_{hidden}^2 =\frac{g^2_0}{Re(S-\alpha^{(2)}T +\b\frac{{\bf Y}^2}{T})}
\label{2.6.2}
\end{equation}
respectively. Here $\alpha^{(1)}$ is the tension of the visible brane
\begin{equation}
\alpha^{(1)} =
\frac{\pi \rho}{16 \pi v_{CY}}
(\frac{\kappa_{11}}{4 \pi})^{2/3}
\int_{CY} \omega \wedge (tr F^{(1)} \wedge F^{(1)} - \frac{1}{2} tr {\cal R}
\wedge {\cal R}),
\label{2.6.3}
\end{equation}
where $F^{(1)}$ is the curvature of the gauge bundle on the visible brane,
and $g^2_0$ is a moduli independent parameter of order $\alpha_{GUT}$.

Now let us estimate the order of magnitude of the parameter $c$.
The axion transformation law is inherited from the variation of
the $B$-field in ten dimensions (or the $C$-field in eleven
dimensions). In the case of strongly coupled heterotic string it
is of the form~\cite{HorWit2}
\begin{equation}
\delta B \sim \frac{1}{4 \sqrt{2} \pi^2 \rho} (\frac{\kappa_{11}}{4 \pi})^{2/3}
(\Omega_{YM} -\frac{1}{2}\Omega_{L}),
\label{2.7}
\end{equation}
where $\Omega_{YM}$ and $\Omega_{L}$ are the usual Chern-Simons forms.
By using eq.~\eqref{1.5.4}, we get
\begin{equation}
c \sim \frac{1}{(4 \pi)^{5/3}} (\frac{1}{M^2_{Pl}\pi \rho})^{4/3}.
\label{2.8}
\end{equation}
Now we can write the expressions for $U_{D}$. If the anomalous $U(1)$ is in the visible sector, we get
\begin{equation}
U_D = M^4_{Pl} \frac{b}{V^2 Re(S+\alpha^{(1)}T +\b (T-\frac{{\bf Y}^2}{T}))},
\label{2.9}
\end{equation}
whereas if the anomalous $U(1)$ appeared in the hidden sector we get
\begin{equation}
U_D = M^4_{Pl} \frac{b}{V^2 Re(S-\alpha^{(2)}T +\b\frac{{\bf Y}^2}{T})}.
\label{2.10}
\end{equation}
The coefficient $b$ is given by
\begin{equation}
b \sim \frac{g_0^2}{(4 \pi)^{10/3}}(\frac{1}{M^2_{Pl}\pi \rho})^{8/3} \sim 10^{-18}.
\label{2.11}
\end{equation}
We see that generically $b \sim \frac{W^2}{M^2_{Pl}}$. This means
that $U_{D}$ is of the same order of magnitude as the supergravity
potential energy. This, in turn, means that it is potentially
possible to obtain a vacuum with very small cosmological constant
by fine tuning.

To conclude this section, let us make some remarks on the charged matter fields. In principle,
$U_D$ contains terms involving the charged matter fields. Giving them generic vacuum expectation values
breaks the low-energy gauge group down to nothing. Therefore we will set them to zero and concentrate
on the moduli potential. We will not discuss potentials for the charged matter fields in this paper.


\section{dS Vacua}


In this section, we will show that the AdS vacuum constructed in Section 2 can indeed be raised to a dS vacuum.
We will see that two of the equations of motion can be solved by balancing the run away potentials
against fluxes.
The remaining equation will be purely algebraic and will be solved numerically.
Despite the fact that the potential energy is a  rather complicated function of three variables,
almost all calculations can be performed analytically.

The potential energy of the system under study is given by
\begin{equation}
\frac{U}{M^4_{Pl}}=U_0 +U_D,
\label{3.1}
\end{equation}
where $U_0$ is the supergravity contribution
\begin{equation}
U_0 = e^{K}(G^{-1}|DW|^2 -3 W \bar W)
\label{3.2}
\end{equation}
and $U_D$ is the contribution coming from Fayet-Iliopoulos terms.
For concreteness, we assume that the anomalous $U(1)$ is in the
visible sector. In this case we have
\begin{equation}
\label{3.3}
U_D =  \frac{b}{V^2 Re(S+\alpha^{(1)}T +\b (T-\frac{{\bf Y}^2}{T}))}.
\end{equation}
To simplify our notation, we have redefined
\begin{equation}
\frac{K}{M^2_{Pl}} \rightarrow K, \quad \frac{W}{M^3_{Pl}} \rightarrow W.
\label{3.4}
\end{equation}
We will argue that this potential energy can admit a minimum with
a positive cosmological constant. We will show that the order of
magnitude of the Kahler covariant derivatives in the minimum is
sufficiently less than one. Therefore, the supergravity
contribution to the vacuum energy is still negative. This leads a
possibility of obtaining a dS vacuum with a small cosmological
constant.

As in Section 2, for simplicity, we will ignore all the imaginary
parts of the fields $S, T$ and ${\bf Y}$. To stabilize them one
can invoke similar arguments as in the case of the AdS
stabilization \cite{BO}. In particular, it is possible to show
that we still have
\begin{equation}
Im(T) \sim \frac{1}{\tau} \approx 0, \quad Im({\bf Y}) \approx 0.
\label{3.5}
\end{equation}
Without loss of generality we can assume that both $W_{g}$ and $W_{np}$ are out of phase with respect
to $W_f$, so that the superpotential for this system is still given by~\eqref{1.5.6}.
Furthermore, since $U_D$ does not depend on the complex structure moduli $Z_{\alpha}$, it is natural to treat
$W_f$ as a constant with all the moduli $Z_{\alpha}$ frozen at values solving eqs.~\eqref{1.20}.
Thus, we can think of $U$ as of a function of three variables $V, R$ and $y$. We cannot assume that any
of these moduli are frozen near the old AdS minimum since $U_D$ depends on all three of them.
To simplify our calculation, we will also assume that the five-brane
Kahler potential (the last term in eq.~\eqref{1.5.2}) is sufficiently
less than the $S$- and $T$-Kahler potentials (the first two terms in eq.~\eqref{1.5.2}),
so that we can ignore the off-diagonal components of the inverse Kahler metric.
The necessary conditions for this is
\begin{equation}
2 \tau_5 \frac{y^2}{VR} <<1.
\label{3.5.1}
\end{equation}
To satisfy it, we will search for a solution with $V \approx 2 -
3$ and $\tau_5 \approx 0.5$. To be able to do this, we will
assume that $W_f$ is of order $10^{-10}$ or less (in $M_{Pl}^3$
units). We found that the easiest way to analyze the equations of
motion is to observe how the Kahler covariant derivatives get
disturbed by the presence of the Fayet-Iliopoulos contribution.
Therefore, we define
\begin{eqnarray}
&& A \equiv D_{S}W = \e W_g -\frac{1}{2V} W_f, \nonumber \\
&& B \equiv D_{T}W = \tau W_{np} -
\e (\alpha^{(2)} + \frac{\b y^2}{R^2})W_g - \frac{3}{2R} W_f, \nonumber \\
&& C \equiv D_{{\bf Y}}W =  -\tau W_{np} + \frac{2 \e \b y}{R}
W_g + \frac{2 \tau_5 y}{VR} W_f,
\label{3.6}
\end{eqnarray}
where eqs.~\eqref{1.5.2}, \eqref{1.5.6}, \eqref{1.7}, \eqref{1.15} and \eqref{1.19} have been used.
As in Section 2 (see eqs.~\eqref{1.21} and~\eqref{1.23}), we will be working in the regime
\begin{equation}
2 \e W_g \sim \tau W_{np} \sim W_f.
\label{3.7}
\end{equation}
In order to be able to fine the cosmological constant we have to have
\begin{equation}
W_f \sim \sqrt{b}.
\label{3.8}
\end{equation}
It was argued in the previous section that this is generically the case.

\begin{figure}
\epsfxsize=3.5in \epsffile{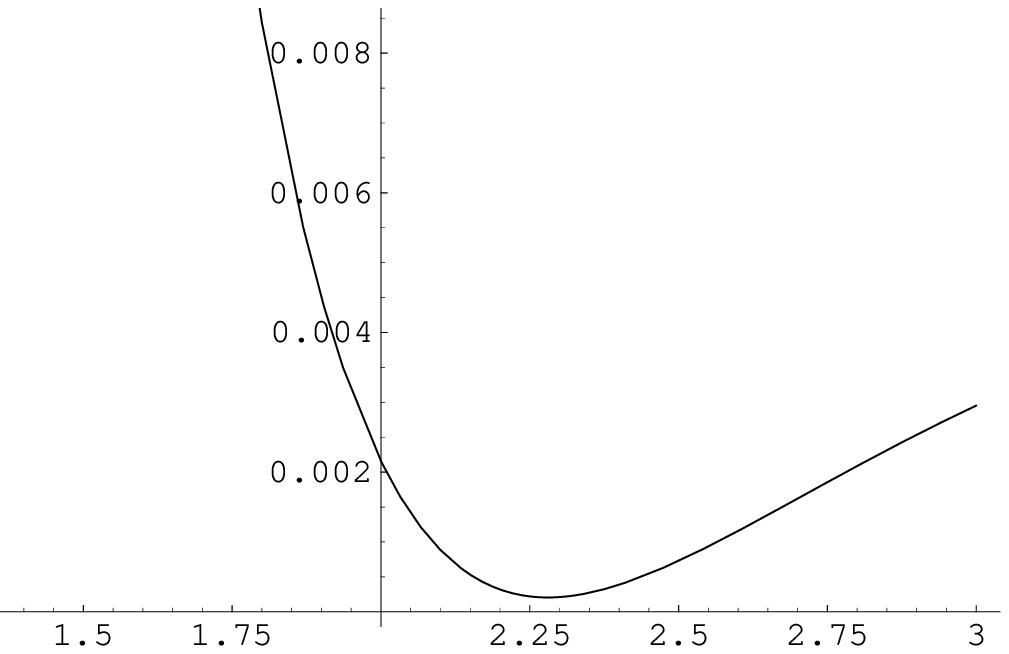}
\begin{picture}(30,30)
\put(-153,160){U}
\put(5, 15){V}
\end{picture}
\caption{A schematic slice of the potential near the dS minimum
(multiplied by $10^{12}$) in the $V$ direction. \label{v1}}
\vspace{0.5cm}
\epsfxsize=3.5in \epsffile{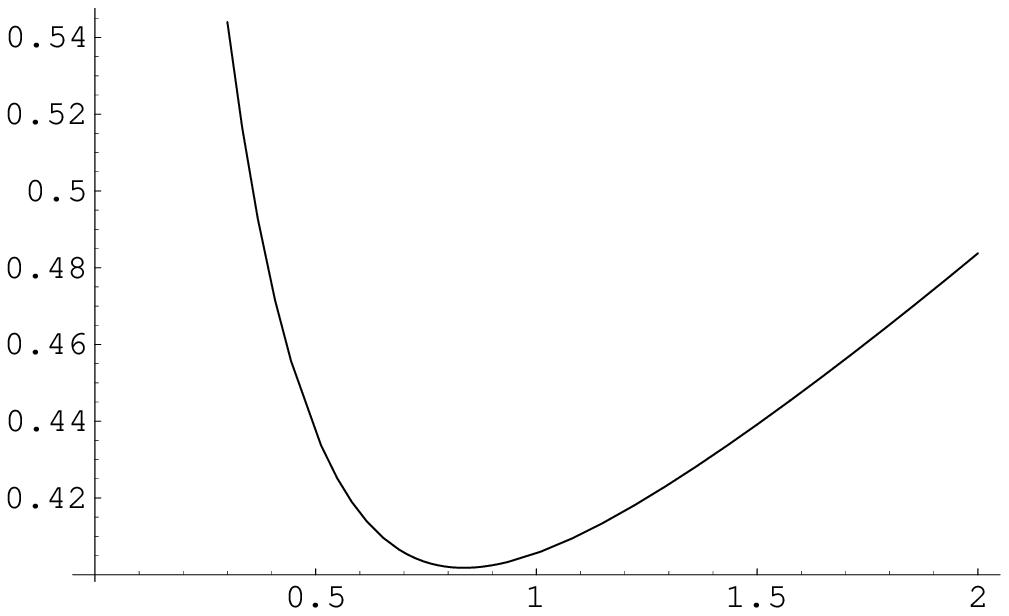}
\begin{picture}(30,30)
\put(-225,150){U}
\put(10, 15){R}
\end{picture}
\caption{A schematic slice of the potential near the dS minimum
(multiplied by $10^{12}$) in the $R$ direction. \label{r1}}
\vspace{0.5cm}
\epsfxsize=3.3in \epsffile{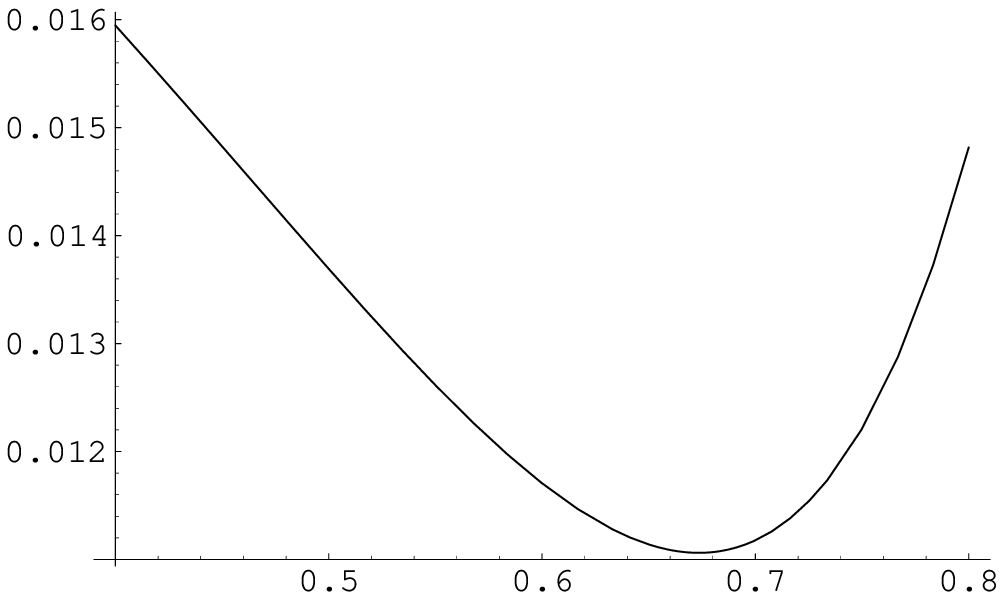}
\begin{picture}(30,30)
\put(-210,140){U}
\put(10, 15){y}
\end{picture}
\caption{A schematic slice of the potential near the dS minimum
(multiplied by $10^{12}$) in the $y$ direction. \label{y1}}
\end{figure}

Before we write the equations of motion, let us take a brief look at the structure
of the derivatives of the supergravity contribution to the potential energy $U_0$.
When we differentiate $U_0$ we will obtain two kinds of terms.
Some terms will involve derivatives of $A, B$ and $C$. We will call such terms ``leading''.
The other terms we will call ``subleading''. It is easy to see
that the ``leading'' terms contain extra factors
of $2\e$ and $\tau$ comparing to the ``subleading'' ones. This means that the ``leading'' terms have a higher
order of magnitude. Clearly, in the equations of motion, it is enough to keep only the ``leading'' terms
(unless they cancel out). First, consider the equation
\begin{equation}
\frac{\partial U}{\partial T}=0.
\label{3.8.1}
\end{equation}
The ``leading'' terms look as follows
\begin{eqnarray}
&&\frac{\partial U_0}{\partial T} =e^{K}
[(2 \e V)^2 (\alpha^{(2)} +\frac{\b y^2}{R^2})
W_g A -\frac{4}{3} R^2 (\tau^2 W_{np} + \e^2 (\alpha^{(2)}+ \frac{\b y^2}{R^2})W_g)B
+ \nonumber \\
&&\frac{V R}{\tau_5}(\tau^2 W_{np} + \frac{2 \e^2 \b
y}{R}(\alpha^{(2)} +\frac{\b y^2}{R^2})W_g)C ]. \label{3.9}
\end{eqnarray}
Recall that by $A, B$ and $C$ we have denoted the Kahler covariant derivatives \eqref{3.6}.
In eq.~\eqref{3.7}, the two terms involving $\tau^2 W_{np}$ are greater than the other terms
by a factor of $\frac{\tau}{2 \e} \sim 20$. This means that
\begin{equation}
\frac{\partial U_0}{\partial T} \approx e^{K}(\frac{VR}{\tau_5} C-\frac{4}{3} R^2 B)\tau^2 W_{np}.
\label{3.10}
\end{equation}
Furthermore,
\begin{equation}
\frac{\partial U_D}{\partial T} =-
\frac{b (\alpha^{(1)} + \b + \frac{by^2}{R^2})}{2V^2 (V+ \alpha^{(1)} R +\b R-\b \frac{y^2}{R})} \equiv -b_T.
\label{3.11}
\end{equation}
From eq.~\eqref{3.7} and \eqref{3.9} it follows that in the interesting range of the fields
\begin{equation}
\frac{\partial U_0}{\partial T} \sim \tau \frac{\partial U_D}{\partial T} >> \frac{\partial U_D}{\partial T}.
\label{3.11.1}
\end{equation}
Therefore, in the minimum the Kahler covariant derivatives $B$ and $C$ are related to each other as follows
\begin{equation}
\frac{V}{\tau_5} C \approx \frac{4R}{3} B.
\label{3.12}
\end{equation}
Now let us consider the equation
\begin{equation}
\frac{\partial U}{\partial S} =0.
\label{3.12.1}
\end{equation}
By using eqs.~\eqref{3.3}, \eqref{3.6} and \eqref{3.12}
and keeping the ``leading'' terms in $\frac{\partial U_0}{\partial S}$,
we get
\begin{equation}
e^{K}[-(2\e V)^2 W_g A+ (\frac{\e^2 VR}{\tau_5} (\alpha^{(2)} +\frac{\b y^2}{R^2})-\frac{2 \e^2 Vy}{\tau_5}) W_g C]-b_S =0,
\label{3.13}
\end{equation}
where
\begin{equation}
b_S =-\frac{\partial U_D}{\partial S} =
\frac{b(\alpha^{(1)}+ \b(1+ \frac{y^2}{R^2}))}{2V^2(V+\alpha^{(1)} R +\b(R-\frac{y^2}{R}))^2}.
\label{3.14}
\end{equation}
By using eqs.~\eqref{3.6} and \eqref{3.12}, eq. \eqref{3.13} can be written as follows
\begin{equation}
-\mu W_g^2 +2 \nu W_f W_g  -b_S e^{-K} =0,
\label{3.15}
\end{equation}
where
\begin{equation}
\mu = V \e^3 \left[ 4V + \frac{((\alpha^{(2)}+\frac{\b y^2}{R^2}) -2 \b y)
((\alpha^{(2)}+\frac{\b y^2}{R^2}) -2 \b y+ 2 \e V (1-\frac{\tau_5}{b}))}{R\tau_5 +\frac{3V}{4}} \right]
\label{3.16}
\end{equation}
and
\begin{equation}
\nu =V\e^2 \left[1 +\frac{(\alpha^{(2)}+\frac{\b y^2}{R^2} -2 \b y)(\frac{8\b y}{V}-2)}{\frac{4R}{3\tau_5}+V}\right].
\label{3.17}
\end{equation}
By using eq.~\eqref{1.7.4}, it is easy to see that $\mu >0$. Without loss of generality we can assume
that $\nu$ is greater than zero. The sign of $\nu$ depends on the relative phase of $W_g$ and $W_f$
which in turn depends on the imaginary parts of the moduli.
Without loss of generality we can assume that $\nu >0$ and the imaginary parts of the moduli are
stabilized in such a way that $W_g$ and $W_f$ are out of phase.
One can show that without the simplifying assumption~\eqref{3.5.1}, the structure of
eq.~\eqref{3.15} would be exactly the same but the coefficients $\mu$ and $\nu$ would
be much more complicated.
If
$b_S e^{-K}$ is not very large, this equation has two solutions for $W_g$ and, hence, for $V$.
It is easy to realize that the smaller solution is the minimum of $U$ and the bigger one is the maximum of $U$.
In the minimum $W_g$ is given by
\begin{equation}
W_{g min} =\frac{1}{\mu} \left( \nu W_f - \sqrt{(\nu W_f)^2 - \mu b_S e^{-K}} \right).
\label{3.18}
\end{equation}
From eqs.~\eqref{3.8}, \eqref{3.16} and \eqref{3.17} it follows that
\begin{equation}
(\nu W_f) \sim \e^4 W_f^2,
\label{3.19}
\end{equation}
whereas
\begin{equation}
\mu b_S \sim \e^3 W_f^2.
\label{3.20}
\end{equation}
Therefore, for interesting relative values of $W_f$ and $b$, the discriminant of the
quadratic (with respect to $W_g$) equation \eqref{3.15} is positive that guarantees the existence
of the minimum. It is clear that if $b$ is relatively large comparing to $W_f$, the minimum disappears.
Eq.~\eqref{3.18} is the analogue of eq.~\eqref{1.21} in Section 2. It stabilizes $V$ at the value of order one
(provided $R$ and $y$ are stabilized). Now we go back to eq.~\eqref{3.12}.
By using eqs.~\eqref{3.6}, it can be written as
\begin{equation}
(\frac{4R^2}{3} +\frac{VR}{\tau_5}) \tau e^{-\tau (R-y)} =\e
(\frac{2  \b y V}{\tau_5} +\frac{4R^2}{3}  (\alpha^{(2)}+\frac{\b y^2}{R^2})) W_g+
2(R+y)W_f,
\label{3.21}
\end{equation}
where $W_g$ is given by eq.~\eqref{3.18}. This equation is the analogue of eq.~\eqref{1.23}
in Section 2. As it was discussed before, we obtain
\begin{equation}
R-y \approx 0.1.
\label{3.22}
\end{equation}
Now we consider the last equation
\begin{equation}
\frac{\partial U}{\partial{\bf Y}} =0.
\label{3.23}
\end{equation}
If, as before, we take only the ``leading'' terms in $\frac{\partial U_0}{\partial{\bf Y}}$,
they turn out to be a linear combination of the ``leading'' terms of the equations
$\frac{\partial U_0}{\partial S}$ and $\frac{\partial U_0}{\partial T}$. Therefore, we have to
include all the ``subleading'' terms. After tedious calculations one can derive the following equation
\begin{eqnarray}
&& (-2AW_f -\frac{y}{V} W_f C +2 V A^2 -\frac{2 R^2}{3V} B^2)
(\alpha^{(2)} +\frac{\b y^2}{R^2} -\frac{2 \b y}{R}) - \nonumber \\
&& 2BW_f -\frac{y}{R}W_f C -\frac{2R}{3}B^2 -\frac{V}{\b} C^2 - 2 C W_f +
\frac{8 \tau_5 y}{R V} A^2 + \nonumber \\
&& \frac{8 \tau_5 y}{3 R V} B^2 + 2 y C^2 = [b_S(\alpha^{(2)}
+\frac{\b y^2}{R^2} -\frac{2 \b y}{R}) +b_T - b_{\bf Y}]e^{-K}.
\label{3.24}
\end{eqnarray}
In this equation, $A, B$ and $C$ are Kahler covariant derivatives \eqref{3.6}, $W_g$ is given by
eq.~\eqref{3.18}, \eqref{3.17} and \eqref{3.16}, $C$ and $B$ are related by eq.~\eqref{3.12},
$b_S$ and $b_T$ are given by eqs.~\eqref{3.14} and \eqref{3.11} respectively and
$b_{{\bf Y}}$ is given by
\begin{equation}
b_{{\bf Y}} = \frac{b \b y}{V^2 R (V+\alpha^{(1)}R + \b (R-\frac{y^2}{R})^2}.
\label{3.25}
\end{equation}
Eq.~\eqref{3.24} is the analogue of eq.~\eqref{1.26} in Section 2.
A numeric analysis shows that it is possible to find a solution
to eq.~\eqref{3.24} satisfying all the right conditions. In order
to justify our neglecting the off-diagonal components of the
inverse Kahler metric, we have to take $V \approx 2-3$. We also
would like to prove that it is conceivable to fine tune the
cosmological constant to zero. For this to be true, it is
necessary to remain the supergravity contribution to the
cosmological constant $U_{0 min}$ negative. If we take
\begin{equation}
\b = 0.5, \quad \tau_5 =0.5, \quad \alpha^{(1)} =0.5, \quad {\alpha^{(2)}} =1, \quad R \approx 0.8,
\quad V \approx 2.3, \quad \frac{be^{-K}}{W_{f}} \approx 4.2,
\label{3.26}
\end{equation}
we find that there is a unique positive solution for $y$
\begin{equation}
y \approx 0.7.
\label{3.27}
\end{equation}
The Kahler covariant derivatives (multiplied by the corresponding components of the inverse Kahler metric)
are given by
\begin{equation}
G^{-1}_{S \bar S} A^2 \approx 0.7 W_f^2, \quad G^{-1}_{T\bar T}B^2 \approx 0.4 W_f^2, \quad
G^{-1}_{{\bf Y} \bar {\bf Y}} C^2 \approx 0.2 W_f^2.
\label{3.28}
\end{equation}
This means that the supergravity contribution to the cosmological constant is negative.
Various slices of the potential energy near the minimum are
schematically shown on Figures~\ref{v1}-\ref{y1}. Clearly, since
the supergravity contribution to the cosmological constant is
negative, by fine tuning the ratio $\frac{b}{W_{f}^2}$, it is
possible to obtain the cosmological constant of order of the
experimentally observed value
\begin{equation}
\Lambda \sim 10^{-120}M_{Pl}^4.
\label{3.29}
\end{equation}
However, note that if the number of moduli is sufficiently big, it is very likely that the supergravity
contribution to the potential energy will become positive. This means that even though it is still possible
to stabilize the moduli in a dS minimum, the value of the cosmological constant cannot be
made small even by fine tuning.

This concludes our analysis. We have shown that by using
Fayet-Iliopoulos terms, it is conceivable to obtain a vacuum with
a positive cosmological constant which can be fine tuned to the
experimentally observed value. In this section, we have used the
expression for the Fayet-Ilipoulos contribution to the potential
energy given by eq.~\eqref{2.9}. That is, we assumed that the
anomalous $U(1)$ was in the visible sector. Clearly, with the
same success we could assume that the anomalous $U(1)$ was in the
hidden sector and use the Fayet-Iliopoulos contribution to the
potential energy given by eq.~\eqref{2.10}. Let us make a brief
remark on the vector bundle moduli. It seems straightforward to
add them to our analysis. Since the Fayet-Iliopoulos contributions
do not depend on them, one can argue that they will be frozen
roughly at the same values as in the old AdS vacuum. No
conceptually new difficulties are expected.


\section{Moduli Potentials in the $E_8 \times \bar E_8$ Theory}


In this section, we consider what kind of moduli potentials in the low-energy field theory we get
if we break supersymmetry in the bulk. This can be achieved either by adding antibranes or
by a chirality change at one of the orbifold fixed planes. The theory obtained in the
latter case was called the $E_8 \times \bar E_8$ theory in \cite{Fabinger}.
In both cases, the functional form of the moduli
potential is the same.
We will concentrate mostly on the $E_8 \times \bar E_8$ theory.
We will assume that the fermions on the visible brane have
positive chirality, whereas the fermions on the hidden brane have
negative chirality. We will refer to the sector on the visible
brane as to the $E_8$ sector and to the sector on the hidden
brane as to the $\bar E_8$ sector. The $E_8 \times \bar E_8$
theory is, clearly, non-supersymmetric. In the efffective field
theory, supersymmetry is broken explicitly by the gravitino mass.
Indeed, the $E_8 \times \bar E_8$ theory can be viewed as the
eleven-dimensional supergravity on $S^1/Z_2$ with the gravitino
antiperiodic along the circle \cite{Fabinger}. This is equivalent
to saying that, in the effective field theory, the gravitino has
a mass. On the other hand, in the gauge theory sector,
supersymmetry is not broken, at least at the tree level. This
implies that when we compactify the theory on a Calabi-Yau
threefold to obtain the effective theory in four-dimensions,
supersymmetry will be broken in the gravity sector and preserved
in the gauge and the matter sectors. Despite the fact that in the
effective field theory the gravitino is massive, the theory still
might suffer from local anomalies. They are cancelled by
arguments similar to ones discussed in \cite{HorWit2}. Let us
briefly go through them. As before, we will denote the gauge
field strength on the visible brane by $F^{(1)}$ and on the
hidden brane by $F^{(2)}$. In the $E_8$ sector, the anomaly is
described \cite{HorWit2} by the twelve-form
\begin{equation}
I_{12} = I_4 \wedge I_8,
\label{4.1}
\end{equation}
where
\begin{equation}
I_4 = \frac{1}{2} tr{\cal R} \wedge {\cal R} -tr F^{(1)} \wedge F^{(1)}
\label{4.2.1}
\end{equation}
and
\begin{equation}
I_8 =-\frac{1}{4} trI_4^2 + I^{\prime}_8,
\label{4.2.2}
\end{equation}
where
\begin{equation}
I^{\prime}_8 = -\frac{1}{4} I_4^2 +[-\frac{1}{8}tr {\cal R}^4 +\frac{1}{32} (tr {\cal R} \wedge {\cal R})^2].
\label{4.3}
\end{equation}
Locally $I_{12}$ can be written as
\begin{equation}
I_{12} = d(I_3) \wedge I_8 = d(I_3 \wedge I_8),
\label{4.4}
\end{equation}
with $I_3$ being the difference of the Chern-Simons forms
\begin{equation}
I_3 =\frac{1}{2}\Omega_3 (\omega_L) -\Omega_3(A^{(1)}),
\label{4.5}
\end{equation}
where $\omega_L$ is the spin connection. Under gauge and locally Lorentz transformations the
polynomial $I_3 \wedge I_8$ transforms as follows
\begin{equation}
\delta (I_3 \wedge I_8) = d(I_2 \wedge I_8),
\label{4.6}
\end{equation}
where
\begin{equation}
I_2 = \frac{1}{2} tr(\theta {\cal R}) - tr(\e F^{(1)}).
\label{4.7}
\end{equation}
This allows us to conclude that the variation of the effective action on the visible brane
is given by
\begin {equation}
(\delta \Gamma)|_{x^{11}=0} \sim \int  (I_2\wedge I_8) =
\int d^{10} x (\frac{1}{2} tr(\theta {\cal R}) - tr(\e F^{(1)})) \wedge I_8.
\label{4.8}
\end{equation}
To cancel this anomaly, it necessary to modify the Bianchi identity for the $G$-flux \cite{HorWit2}
\begin{equation}
(dG)|_{x^{11}=0} \sim \frac{1}{2} tr {\cal R}\wedge {\cal R} - tr F^{(1)} \wedge F^{(1)}.
\label{4.9}
\end{equation}
This implies that the three-form potential $C$ is not gauge invariant and transforms as follows
\begin{equation}
(\delta C)|_{x^{11}=0} \sim -\frac{1}{2} tr (\theta {\cal R}) +\tr (\e F^{(1)}).
\label{4.10}
\end{equation}
The anomaly~\eqref{4.8}
is cancelled by the Chern-Simons coupling of the eleven-dimensional supergravity
\begin{equation}
\Gamma_{CS} \sim \int C\wedge G \wedge G,
\label{4.11}
\end{equation}
as well as proposed Green-Schwarz interaction
\begin{equation}
\Gamma_{GS} \sim C \wedge I^{\prime \prime}_8.
\label{4.12}
\end{equation}
Here $I^{\prime \prime}_8$ is given by
\begin{equation}
I^{\prime \prime}_8 =
- \frac{1}{4} (\frac{1}{2} tr {\cal R} \wedge {\cal R} -
tr F^{(1)} \wedge F^{(1)} -
tr F^{(2)} \wedge  F^{(2)})^2
+[-\frac{1}{8}tr {\cal R}^4 +\frac{1}{32} (tr {\cal R} \wedge {\cal R})^2].
\label{4.12.1}
\end{equation}
On the visible brane we have
\begin{equation}
I^{\prime \prime}_8 =I^{\prime}_8.
\label{4.13.1}
\end{equation}
The explicit proportionality coefficients in eqs.~\eqref{4.8}-\eqref{4.12}
are not important here and can be found in~\cite{HorWit2}.
It is straightforward to see that
\begin{equation}
(\delta \Gamma)|_{x^{11}=0} +\delta (\Gamma_{CS} +\Gamma_{GS})|_{x^{11}=0}=0.
\label{4.13}
\end{equation}
In the $\bar E_8$ sector, the anomaly consideration is analogous. Due to a chirality change, the anomaly
polynomial $\bar I_{12}$ is given by
\begin{equation}
\bar I_{12} =\bar I_4 \wedge \bar I_8,
\label{4.14}
\end{equation}
where
\begin{equation}
\bar I_4= -(\frac{1}{2} tr {\cal R} \wedge {\cal R} -tr F^{(2)} \wedge  F^{(2)})
\label{4.15}
\end{equation}
and
\begin{equation}
\bar I_{8} =-\frac{1}{4}  \bar I_4^2 +   \bar I^{\prime}_8,
\label{4.16}
\end{equation}
where
\begin{equation}
\bar I^{\prime}_8 = -\frac{1}{4} \bar I_4^2 +[-\frac{1}{8}tr {\cal R}^4
+\frac{1}{32} (tr {\cal R} \wedge {\cal R})^2].
\label{4.16.1}
\end{equation}
The variation of the effective action on the hidden brane
is given by the expression similar to eq.~\eqref{4.8}
\begin{equation}
(\delta \Gamma)|_{x^{11}=\pi \rho} \sim
\int d^{10} x (-\frac{1}{2} tr(\theta {\cal R}) +tr(\e F^{(2)})) \wedge I_8.
\label{4.17}
\end{equation}
The modified Bianchi identities and the transformation law of the three-form field $C$ become
\begin{equation}
(dG)|_{x^{11}=\pi \rho} \sim -\frac{1}{2} tr {\cal R}\wedge {\cal R} + tr F^{(2)} \wedge F^{(2)}
\label{4.18}
\end{equation}
%
and
\begin{equation}
(\delta C)|_{x^{11}=\pi \rho} \sim \frac{1}{2} tr (\theta {\cal R}) -\tr (\e F^{(2)}).
\label{4.19}
\end{equation}
By using eqs.~\eqref{4.11}-\eqref{4.12.1}, \eqref{4.17} and \eqref{4.19}, it is straightforward to see that
\begin{equation}
(\delta \Gamma)|_{x^{11}=\pi \rho } +\delta (\Gamma_{CS} +\Gamma_{GS})|_{x^{11}=\pi \rho}=0.
\label{4.20}
\end{equation}
Therefore, the local anomaly cancels in the $\bar E_8$ sector as well.
From the modified Bianchi identities~\eqref{4.9} and~\eqref{4.18}, it follows that if we consider a Calabi-Yau
compactification to four dimensions, the anomaly cancellation condition reads
\begin{equation}
c_2(V_1) =c_2(V_2),
\label{4.21}
\end{equation}
where by $V_1$ and $V_2$ we denoted the gauge bundles on the visible and on the hidden branes respectively.
This condition gets modified by the presence of (anti) five-branes (wrapped on holomorphic cycles) in the bulk.
Recall \cite{Duff}, that every five-brane contributes to the anomaly as one instanton in the
$E_8$ sector and every anti five -brane contributes as one instanton in the $\bar E_8$ sector.
Hence, the modified
anomaly cancellation condition reads
\begin{equation}
c_2(V_1)-c_2(V_2) +[{\cal W}] -[\bar {\cal W}]=0,
\label{4.22}
\end{equation}
where the last two terms represent the five-brane and the anti five-brane classes respectively.
For completeness and further reference it is useful to recall that, in the standard
$E_8 \times E_8$ theory with
(anti) five-branes, the anomaly cancellation condition is given by \cite{Witten96}
\begin{equation}
c_2(V_1)+ c_2(V_2)-c_2(TX) +[{\cal W}] -[\bar {\cal W}]=0.
\label{4.22.1}
\end{equation}

The moduli potential in the four-dimensional effective field theory will come from
the boundary terms of the eleven-dimensional supergravity
\begin{eqnarray}
S_{boundary} & = & -\frac{1}{8 \pi \kappa_{11}^2} (\frac{\kappa_{11}}{4 \pi})^{2/3}
\int d^{10}x \sqrt{-G_{10}} (tr (F^{(1)})^2 -\frac{1}{2} tr {\cal R}^2)  \nonumber \\
 & & -\frac{1}{8 \pi \kappa_{11}^2} (\frac{\kappa_{11}}{4 \pi})^{2/3}
\int d^{10}x \sqrt{-G_{10}} (tr (F^{(2)})^2 -\frac{1}{2} tr {\cal R}^2).
\label{4.23}
\end{eqnarray}
Upon the compactification to four-dimensions the metric is written as follows~\cite{LOW4}
\begin{equation}
ds_{11}^2=R^{-1}V^{-2/3} g_{4 \mu \nu}dx^{\mu} dx^{\nu} + V^{1/3} g_{CY AB}dx^{A}dx^{B} + V^{-2/3}R^2 (dx^{11})^2.
\label{4.24}
\end{equation}
This form of the metric guarantees that the Einstein in the action is properly normalized~\cite{LOW4}.
Substituting this metric into eq.~\eqref{4.23} and using the identities
\begin{equation}
\int_{CY} \sqrt{g_{CY}}tr F_{AB}^2= - 2 \int_{CY}\omega \wedge F \wedge F = 32\pi^2
\int_{CY}\omega \wedge c_2(V)
\label{4.25}
\end{equation}
and
\begin{equation}
\int_{CY} \sqrt{g_{CY}}tr {\cal R}_{AB}^2= - 2 \int_{CY}\omega \wedge {\cal R} \wedge {\cal R}=
32\pi^2 \int_{CY} \omega \wedge c_2(TX),
\label{4.26}
\end{equation}
for $F$ and $R$ satisfying the Hermitian Yang-Mills equations, we obtain the following
moduli dependent potential
\begin{equation}
\Delta U =\frac{8 \pi}{\kappa_{11}^2}(\frac{\kappa_{11}}{4 \pi})^{2/3}
\int_{CY} \omega \wedge (c_2(V_1) +c_2(V_2) -c_2(TX))
\frac{1}{V R^2}.
\label{4.27}
\end{equation}
%
In the absence of fve-branes in the bulk, in the $E_8 \times E_8$ theory this potential vanishes identically
since the combination
\begin{equation}
J =c_2(V_1)+c_2(V_2)-c_2(TX)
\label{4.28}
\end{equation}
vanishes by the anomaly cancellation condition~\eqref{4.22.1}.
In the presence of five-branes, this potential
cancels against similar terms coming from the five-brane world volume theory (see below).
On the other hand, in the $E_8 \times \bar E_8$ theory, the anomaly cancellation condition looks different.
In the absence of (anti) five-branes it is given by eq.~\eqref{4.21} and the combination~\eqref{4.28}
does not vanish anymore. Terms with the identical functional structure come from the theory on the (anti) five-brane
world-volume. The relevant part of the action of the (anti) five-brane is
given by
\begin{equation}
S_5 =-T_5 \int d^6 \zeta \sqrt{h} + \dots,
\label{4.29}
\end{equation}
where $T_5$ is the (anti) five-brane tension and $h_{rs}$ is the pullback of the
eleven-dimensional metric~\eqref{4.24}
\begin{equation}
h_{rs}=\frac{\partial x^{M}}{\partial \zeta^{r}} \frac{\partial x^{N}}{\partial \zeta^{s}} G_{MN}.
\label{4.30}
\end{equation}
Now we impose the gauge
\begin{equation}
x^{\mu} =\zeta^{\mu}, \quad \mu =0, 1, 2, 3,
\label{4.31}
\end{equation}
and identify $\frac{x^{11}}{\pi \rho}$ with the real part of five-brane modulus $y$.
Then, by using eq.~\eqref{4.24}, the components of the induced metric $h_{rs}$ can be written as follows
\begin{equation}
h_{\mu \nu} = V^{-2/3}R^{-1}g_{4 \mu \nu} + V^{-2/3} (\pi \rho R)^2 \partial_{\mu}y \partial_{\nu}y
\label{4.32}
\end{equation}
and
\begin{equation}
h_{\sigma \tau} = \frac{\partial x^{A}}{\partial \zeta^{\sigma}}
\frac{\partial x^{B}}{\partial \zeta^{\tau}} V^{1/3} g_{CY AB}, \quad \sigma, \tau =1,2.
\label{4.33}
\end{equation}
The metric $h_{\sigma \tau}$ is just the induced metric on the
holomorphic curve on which the (anti) five-brane is wrapped.
Performing the integration over the area of the holomorphic curve
in~\eqref{4.29}, we obtain a potential of the functional
form~\eqref{4.27}. As we mentioned before, in the $E_8 \times
E_8$ compactification, the potential~\eqref{4.27} vanishes,
whereas in the $E_8 \times \bar E_8$ theory it does not. Let us
also point out that if one takes care of the terms involving two
derivatives of $y$ one obtains the Kahler potential for this
modulus given by the last term in eq.~\eqref{1.5.2}. The moduli
potential~\eqref{4.27} has a similar functional structure as the
Fayet-Iliopoulos potentials \eqref{2.9} and~\eqref{2.10}.
Therefore,  potentially, it can be used to raise the AdS vacuum
discussed in Section 2. However, we will see that it is not very
easy to use this potential to stabilize the moduli. The problem
comes from the overall scale of the potential. The
potential~\eqref{4.27} has the following structure
\begin{equation}
\Delta U = \frac{a}{VR^2},
\label{4.34}
\end{equation}
where the coefficient $a$ is generically of order
\begin{equation}
a \sim
\frac{8 \pi v_{CY}^{1/3}}{\kappa_{11}^2}(\frac{\kappa_{11}}{4 \pi})^{2/3}.
\label{4.35}
\end{equation}
By using eqs.~\eqref{1.4.1} and \eqref{1.5.4}, we can estimate $a$ and obtain
\begin{equation}
a \sim M_{Pl}^4 10^{-10}.
\label{4.36}
\end{equation}
This order of magnitude is too big. For a generic compactification, it does not seem to be possible to balance
$\Delta U$ against the non-perturbative and the flux-induced superpotentials to provide
an interesting moduli stabilization, at least if $h^{1,1}=1$.
Let us now assume that $h^{1,1}$ is greater than one and try to determine whether it is ever
possible to decrease the coefficient $a$. Let us look more carefully at the combination~\eqref{4.28}.
Expanding the Kahler form $\omega$ in the basis of harmonic forms $\{\omega_I\}$
\begin{equation}
\omega = \sum_{I=1}^{h^{1,1}} b_I \omega_I,
\label{4.37}
\end{equation}
where $b_I$ are related to the real parts of the $h^{1,1}$ moduli
(see \cite{LOW5} for details), we can write the integral of $J$
more explicitly as
\begin{eqnarray}
\int_{CY} \omega \wedge J & = &
-\frac{1}{16 \pi^2} \int_{CY}\sum_{I=1}^{h^{1,1}} b_I \omega_I \wedge
(tr F^{(1)} \wedge  F^{(1)} + tr F^{(2)} \wedge  F^{(2)} - tr {\cal R}\wedge {\cal R}) \nonumber \\
 & = & - \frac{1}{16 \pi^2} \sum_{I=1}^{h^{1,1}} b_I v_I \int_{C_I}
(tr F^{(1)} \wedge  F^{(1)} + tr F^{(2)} \wedge  F^{(2)} - tr {\cal R}\wedge {\cal R}) \nonumber \\
 & = & \sum_{I=1}^{h^{1,1}}  b_I v_I \int_{C_I} (c_2(V_1)+c_2(V_2)-c_2(TX)),
\label{4.38}
\end{eqnarray}
where $\{C_I\}$ are the four-cycles Poincare dual to the basis $\{\omega_I\}$ and
$v_I$ is the volume of the two-cycle, Poincare dual to the four-form
\begin{equation}
c_2(V_1)+c_2(V_2)-c_2(TX),
\label{4.38.1}
\end{equation}
measured with respect to the Kahler forms $\omega_I$. Apparently,
the only way to decrease the order of magnitude of $\Delta U$ is
to consider a Calabi-Yau threefold in which there is at least one
very small cycle and to take $c_2(V_1)+c_2(V_2)-c_2(TX)$
localized precisely on such a cycle. Of course, the four-form
that has to be localized on a small cycle does not necessarily
have to be $c_2(V_1)+c_2(V_2)-c_2(TX)$. It does only if there are
no (anti) five-branes in the bulk. In general, in the $E_8 \times
\bar E_8$ theory this four-form is given by
\begin{equation}
J^{\prime} = c_2(V_1)+c_2(V_2)-c_2(TX) + [{\cal W}] +[\bar {\cal W}] =2c_2(V_1) -c_2(TX) + 2[{\cal W}],
\label{4.39}
\end{equation}
where the anomaly cancellation condition~\eqref{4.22} has been
used. Note that $J^{\prime}$ (more precisely $\int_{CY} \omega
\wedge J^{\prime}$) is not necessarily positive. Therefore, the
potential \eqref{4.34} combined with the supergravity one can
generate a phenomenologically attractive dS minimum even if the
number of moduli is large. A potential of the form~\eqref{4.34}
also exist in the $E_8 \times E_8$ theory with anti five-branes
in the bulk. It has the form
\begin{equation}
\Delta \tilde{U} =\frac{8 \pi}{\kappa_{11}^2}(\frac{\kappa_{11}}{4 \pi})^{2/3}
\int_{CY} \omega \wedge J^{\prime \prime}
\frac{1}{V R^2},
\label{4.41}
\end{equation}
where
\begin{equation}
J^{\prime \prime} = c_2(V_1)+c_2(V_2)-c_2(TX) + [{\cal W}] +[\bar {\cal W}] =
2 [\bar {\cal W}],
\label{4.42}
\end{equation}
where the $E_8 \times E_8$ anomaly cancellation condition eq.~\eqref{4.22.1} has been used.
This potential is strictly positive. In any case, this type of
potentials can be used only if we can find a Calabi-Yau manifold
with relatively small 2-cycles. Of course, what we mean by this
is that one has to be able to stabilize at least one of the
$h^{1,1}$ moduli at a very small scale comparing to the
Calabi-Yau scale. We will not attempt to solve this problem in
this paper. Instead, let us speculate on what could be the key to
solving this problem. The non-perturbative superpotentials
discussed in Section 2 have factors depending on complex
structure and vector bundle moduli~\cite{Witten95, Lima1, Lima2}.
We ignored them as they were not important for our purposes. For
certain geometries those factors were explicitly calculated
in~\cite{BDO1, BDO2} and found to be high degree polynomials. If
one manages to show that the values of such factors associated to
different isolated curves in the Calabi-Yau threefold can vary a
lot, it could stabilize various cycles at different scales. The
high degree polynomials might provide a considerable help.
Unfortunately, to study systematically the vector bundle moduli
contribution to non-perturbative superpotentials is a very
challenging problem.


\section{Conclusion}


In this paper, we considered a problem of moduli stabilization in
a metastable dS vacuum in the context of strongly coupled
heterotic string theory. We showed that, as in type IIB theory
\cite{KKLT}, dS vacua can be constructed by adding various
correction to the supergravity potential energy. We studied two
types of such corrections. The first type corrections are
generated by Fayet-Iliopoulos terms \cite{DSW, Burgess}. They
appear if the low-energy gauge group in the visible or in the
hidden sector contains an anomalous $U(1)$ factor. The form of
the moduli potential energy is slightly different depending on in
which sector there is such a factor. This potential is always
positive. We showed that the total potential energy indeed can
have a dS minimum. Moreover, since the supergravity part of the
potential energy can be kept negative, it is possible, by fine
tuning, parameters to get a cosmological constant consistent with
observations. However, if the number of moduli is sufficiently
large, one can expect that both supergravity and Fayet-Iliopoulos
contributions to the potential energy will become positive and it
will not be possible to obtain a small cosmological constant even
by fine tuning. Corrections of the second type can be generated
if we add anti five-branes in the bulk or, more generally,
consider $E_8 \times \bar E_8$ compactifications with both
five-branes and anti five-branes involved. A moduli potential
arises since the net tension of various branes does not cancel
anymore (though the net charge does). This potential can, in
principle, be both positive and negative. It is also more
universal because its existence does not depend on details of the
compactification. Roughly, it has a similar functional form as
the Fayet-Iliopoulos terms. Therefore, it can generate a dS
minimum in a similar way. However, we noticed that, in a generic
compactification, its order of magnitude is not small enough and
not comparable with that of fluxes and non-perturabtive
superpotentials. To make it comparable, it is necessary to learn
how to stabilize some of the Kahler structure moduli at a scale
sufficiently smaller than the Calabi-Yau scale. One can speculate
that this problem is related to understanding of non-exponential
factors in non-perturbative superpotentials. The results of this
paper can be considered as the heterotic version of the type IIB
moduli stabilization mechanism~\cite{KKLT}. However, there are
some new non-trivial elements. Generic heterotic models always
have more types of moduli and, hence, more types of
superpotentials have to be included. Some of them can be trusted
only in a certain range of the moduli space. One more important
feature is that it seems hard to avoid five-branes.


\section{Acknowledgements}


The author would like to thank Vijay Balasubramanian, Melanie Becker,
Axel Krause, Hitoshi Murayama and especially Juan Maldacena and Burt Ovrut
for lots of helpful discussions.
The work is supported by NSF grant PHY-0070928.



\end{document}